# Relativistic quantum Brownian motion

Roumen Tsekov
Department of Physical Chemistry, University of Sofia, 1164 Sofia, Bulgaria

A Wigner-Klein-Kramers equation is proposed, which merges relativistic, quantum and thermo dynamics. The relativistic effect on quantum Brownian motion is studied via the Breit-Fermi Hamiltonian applied into a dissipative Madelung hydrodynamics. A new thermo-quantum Smoluchowski equation is derived, which accounts for the relativistic correction of the Bohm quantum potential.

In the recent years the interest on the relativistic non-equilibrium thermodynamics has grown. Apart from relativistic gas kinetic theories [1, 2], a theory of relativistic Brownian motion is also developed [3]. After Einstein it is well known that the relativistic energy of a particle in vacuum is given by the expression

$$E = \sqrt{m^2 c^4 + c^2 p^2} + U \qquad (1)$$

where $m$ is the particle mass at rest, $c$ is the speed of light, $p$ is the particle momentum and $U$ is a potential. The root in Eq. (1) complicates substantially the way of quantizing the relativistic motion. In the literature there are two important solutions of this problem: the Klein-Gordon and Dirac equations. Their complex mathematical structures and some problems with definition of the probability density make it difficult to extend these equations for description of relativistic quantum Brownian motion. Moreover, it is not clear if the non-relativistic energy $\hat{E} = i\hbar \partial_t$ and momentum $\hat{p} = -i\hbar \nabla$ operators, used in quantizing of Eq. (1), are valid in the relativistic quantum mechanics as well. The present paper proposes an alternative approach via a relativistic Wigner-Klein-Kramers equation. The quantum Brownian motion is described by the Madelung quantum hydrodynamics [4] coupled to the Breit-Fermi Hamiltonian.

According to the relativistic dynamics the stochastic Langevin equation for a classical relativistic particle moving in a classical non-relativistic environment acquires the form

$$\dot{r} = p/M \qquad \dot{p} = -\nabla U - b\dot{r} + f \qquad (2)$$

where $M \equiv \sqrt{m^2 + p^2/c^2}$ is the relativistic dynamic mass, $b$ is the Brownian particle friction coefficient and $f$ is the stochastic Langevin force. It is easy to show that the phase-space probability density $W(p,r,t)$ corresponding to Eq. (2) obeys a relativistic Klein-Kramers equation

$$\partial_t W + p \cdot \nabla W / M - \nabla U \cdot \partial_p W = b \partial_p \cdot (pW / M + k_B T \partial_p W) \tag{3}$$

where $T$ is temperature. The equilibrium solution of Eq. (3) is the Boltzmann-Jüttner distribution $W_{BJ} \sim \exp(-E/k_B T)$ [5], where the energy $E$ is given by Eq. (1). Thus, the classical relativistic Brownian motion is described well by Eq. (2) and the corresponding Eq. (3). Since the relativity affects solely the mass $M$ the overdamped diffusion of a relativistic Brownian particle in the coordinate space will obey the classical non-relativistic Smoluchowski equation, which is independent of the particle mass. A way to describe the relativistic quantum Brownian motion is to extend Eq. (3) to a relativistic Wigner-Klein-Kramers equation

$$\partial_t W + p \cdot \nabla W / M - [U(r + i\hbar \partial_p / 2) - U(r - i\hbar \partial_p / 2)]W / i\hbar = b \partial_p \cdot (pW / M + k_B T \partial_p W) \tag{4}$$

This equation reduces to Eq. (3) in the classical limit $\hbar \to 0$. It represents a complete relativistic extension of the Wigner equation from the high-temperature Caldeira-Leggett approach. The only limitation in Eq. (4) is a classical treatment of the heat bath but most of the interesting systems represent a quantum Brownian particle in a classical environment. However, Eq. (4) does not provide the exact equilibrium Gibbs distribution since a thermo-quantum diffusive operator is missing on the right hand side [6]. Nevertheless, in the case of vacuum ($b = 0$) Eq. (4) reduces to a relativistic Wigner-Liouville equation, which sets a new relativistic quantum mechanics. This alternative to the Klein-Gordon and Dirac formulations does not include four-vectors and deserves alone a separate investigation in the future. For instance, we expect the ground state quantum distribution of a relativistic harmonic oscillator with an own frequency $\omega_0$ to be in the form $W \sim \exp(-2E/\hbar \omega_0)$. Thus, the position dispersion of this oscillator is the same as in the nonrelativistic quantum theory $\sigma_x^2 = \hbar / 2m\omega_0$ while its momentum dispersion is a superposition of nonrelativistic and photon components $\sigma_p^2 = m\hbar \omega_0 / 2 + (\hbar \omega_0 / c)^2 / 2$.

In the case of a relativistic Brownian motion the particle velocity is always much smaller than the speed of light due to friction in the environment. In this case Eq. (1) can be approximated by a truncated momentum power series $E = mc^2 + p^2/2m - p^4/8m^3c^2 + U$. Replacing here the momentum by the corresponding non-relativistic quantum operator $\hat{p} = -i\hbar \nabla$, which is applicable in such semiclassical descriptions, results in the Breit-Fermi Hamiltonian [7]

$$\hat{H}_{BF} = -\hbar^2 \Delta / 2m - \hbar^4 \Delta^2 / 8m^3 c^2 + U \tag{5}$$

where $\Delta \equiv \nabla^2$ is the Laplace operator. Since the relativistic effect is only a correction in Eq. (5), the wave function evolution will obey the standard non-relativistic Schrödinger equation

$$i\hbar\partial_t\psi = \hat{H}_{BF}\psi \tag{6}$$

The quadrate of the wave function modulus is the probability density $\rho \equiv \int Wdp = \bar{\psi}\psi$, which satisfies obligatorily a continuity equation $\partial_t\rho = -\nabla\cdot j$. Employing here the Schrödinger equation (6) provides a new expression for the probability flux $j \equiv \int (p/M)Wdp$

$$j = (i\hbar/2m)[(\psi\nabla\bar{\psi} - \bar{\psi}\nabla\psi) + \lambda_C^2(\nabla\bar{\psi}\Delta\psi - \bar{\psi}\nabla\Delta\psi + \psi\nabla\Delta\bar{\psi} - \nabla\psi\Delta\bar{\psi})] \tag{7}$$

As is seen, there is a relativistic correction of the flux proportional to the square of the particle Compton wavelength $\lambda_C \equiv \hbar/2mc$. For electrons, for instance, this relativistic de Broglie wavelength equals to about 2 mÅ, which corresponds to about 70 electron radii, $r_e = e^2/4\pi\varepsilon_0 mc^2$.

The wave function can be presented in its polar form $\psi = \sqrt{\rho}\exp(iS/\hbar)$ [4]. Since the Brownian motion is relatively slow due to friction we will consider in what follows that the relativistic corrections apply only on the probability density $\rho$. Thus, neglecting the relativistic corrections on the phase $S$, generally associated with the particle velocity [8], Eq. (7) reduces to the standard expression $j = \rho\nabla S/m$. Hence, the continuity equation acquires the form [4]

$$\partial_t\rho = -\nabla\cdot(\rho V) \tag{8}$$

where $V \equiv \nabla S/m$ is the hydrodynamic-like velocity in the probability space. Note that the latter is always smaller than the real velocity of the quantum particle. This is also the case in the classical Brownian motion, where the rate of diffusion is much slower that the thermal motion. The evolution of $V$ can be obtained from Eq. (6) in the frames of the considered strong inequality $V \ll c$

$$m\partial_t V + mV\cdot\nabla V + bV = -\nabla(\rho^{-1/2}\hat{H}_{BF}\rho^{1/2} + k_BT\ln\rho) \tag{9}$$

We introduced here ad hoc a friction force, linear on the velocity, and a thermal entropic force multiplied by the environmental temperature, being the macroscopic image of the Langevin force. Thus Eqs. (8) and (9) represent a dissipative relativistic Madelung quantum hydrodynamics. Note that the present theory describes the relativistic Brownian motion of a single quantum particle in a classical environment; hence no entanglement takes place. Due to the considered slow hydrodynamic motion the relativistic effect in Eq. (9) is purely quantum. Hence, in the classical limit $\hbar \to 0$ Eq. (9) reduces to one for the classical non-relativistic Brownian motion

$$m\partial_t V + mV \cdot \nabla V + bV = -\nabla(U + k_B T \ln\rho) \tag{10}$$

If the friction is very strong one can neglect the first two inertial terms in Eq. (9) to obtain an expression for the hydrodynamic-like velocity, $V = -\nabla(\rho^{-1/2}\hat{H}_{BF}\rho^{1/2} + k_B T \ln\rho)/b$. Substituting it into the continuity equation (8) results in a relativistic thermo-quantum diffusion equation

$$\partial_t \rho = \nabla \cdot [\rho\nabla(\rho^{-1/2}\hat{H}_{BF}\rho^{1/2})/b + D\nabla\rho] \tag{11}$$

where the diffusion constant is given by the classical Einstein expression $D \equiv k_B T/b$. Introducing the Bohm quantum potential $Q[\rho^{1/2}] \equiv -\hbar^2 \Delta\rho^{1/2}/2m\rho^{1/2}$ [8] Eq. (11) can be rewritten as a nonlinear relativistic quantum Smoluchowski equation

$$\partial_t \rho = \nabla \cdot \{\rho\nabla[U + (1 - Q[\rho^{1/2}Q]/2mc^2)Q]/b + D\nabla\rho\} \tag{12}$$

As is seen the relativistic correction of the quantum potential can be expressed by $Q$ as well. This equation could be particularly important for description of relativistic effects in quantum tunneling. In the limit $c \to \infty$ Eq. (12) reduces to a known non-relativistic quantum Smoluchowski equation [9], which relation to the existing theories of non-relativistic quantum Brownian motion is discussed in [10]. As a rule the standard models, reviewed in [11, 12] for instance, concentrate on the effect of a quantum environment. For this reason they predict classical diffusion for a free quantum Brownian particle moving in a classical environment, which violates, however, the Heisenberg principle at short times [13].

In general, the nonlinear Smoluchowski equation (12) is a complex mathematical problem. The relativistic correction leads to violation of the Gaussian character of the free quantum Brownian motion and even for a harmonic oscillator we cannot solve Eq. (12) analytically due to nonlinearity. It is possible, however, to derive a quasi-equilibrium semiclassical approximation of Eq. (12) by replacing the probability density in the quantum potentials via the classical equilibrium Boltzmann distribution $\rho_B \sim \exp(-U/k_B T)$. For instance, using the classical barometric distribution with $U = mgz$ the relativistic correction in Eq. (12) increases of the quantum potential by a factor $(T_g/T)^2$, where $T_g = \hbar g/4ck_B = 6.25 \times 10^{-20}$ K is the gravitational Hawking-Unruh temperature [14, 15]. Furthermore, one should replace $m$ in the Newtonian gravitational potential by the relativistic mass $M$, which results to the Manev semiclassical potential [16] as a first approximation of the Einstein general theory of relativity (see the Appendix). Since the mass of a photon is equal to $M = p/c = 2\pi\hbar/\lambda c$, the Schwarzschild radius $R = 2GM/c^2$ of a photon is larger than its size, equal to the wavelength $\lambda$, if $\lambda < \sqrt{4\pi\hbar G/c^3} \approx 5 \times 10^{-35}$ m. Therefore, pho-tons with wavelength shorter than the Planck length are black holes and should be considered as carriers of the dark energy. Obviously, these black photons can propagate in vacuum only and another their interesting feature is that the light is self-adsorbed.

Alternatively, at high temperatures the quasi-equilibrium semiclassical approximation of Eq. (12) can be linearized for small potentials to obtain

$$\partial_t \rho = \nabla \cdot [\rho \nabla (U + \lambda_T^2 \Delta U + \lambda_C^2 \lambda_T^2 \Delta^2 U)/b + D\nabla \rho] = \nabla \cdot (\rho \nabla U_{eff}/b + D\nabla \rho) \qquad (13)$$

where $\lambda_T \equiv \hbar/2\sqrt{mk_B T}$ is the thermal de Broglie wavelength. If the external potential is a periodic function, e.g. $U \sim \cos(qx)$, the effective potential $U_{eff} = [1-(1-\lambda_C^2 q^2)\lambda_T^2 q^2]U$ is just proportional to the external one. Hence, the quantum effect reduces the energy barriers due to the tunneling effect and the diffusion is free at temperature $T = \lambda_C^2 q^2 (1-\lambda_C^2 q^2)mc^2/k_B$ since $U_{eff} = 0$. The relativistic effect decreases the tunneling effect and, for instance, in structures with $q = 1/\lambda_C$ no tunneling takes place since $U_{eff} = U$. This is, however, expected since the relativity makes the particles heavier, which suppresses the tunneling rate. Another interesting case is a double-well potential $U = -2\alpha x^2 + \beta x^4$, employed widely in bistable systems, with the corresponding effective potential $U_{eff} = -4(\alpha - 6\beta\lambda_C^2)\lambda_T^2 - 2(\alpha - 6\beta\lambda_T^2)x^2 + \beta x^4$. As is seen, the barrier energy $E_a = \beta(\alpha/\beta - 6\lambda_T^2)^2$ is decreased by the quantum tunneling and drops to zero at temperature $T = 3\beta\hbar^2/2\alpha mk_B$; it is not affected, however, by the relativity in this case. It is well-known that the linear non-relativistic quantum Smoluchowski equation from the literature cannot capture deep tunneling. This is true since it represents, in fact, a non-relativistic quasi-equilibrium semiclassical approximation of the nonlinear density-functional quantum Smoluchowski equation (12) [17]. As a result, the linear quantum Smoluchowski equation reduces, similarly to Eq. (13), to the classical diffusion equation in the case of a free quantum Brownian particle ($U = 0$), which obviously violates the Heisenberg principle at short times [13]. The correct full description of the quantum tunneling is described by the nonlinear system of Eqs. (8) and (9) or by Eq. (4). Unfortunately, the mathematics is frustrated by strong nonlinearity.

A dimensional analysis shows that the relative relativistic correction of the quantum potential scales by the ratio of the latter and the energy $mc^2$ at rest. Since $Q$ exhibits huge values at places, where the probability density is extremely low, the conclusion is that important quantum-relativistic corrections are present at these points. This is not surprising since at such places the particle velocity is commensurable with the speed of light [18]. However, the relativistic correction in Eq. (12) decreases the quantum potential there and, hence, it prevents probably the particle to move superluminally. The present relativistic Madelung hydrodynamics should not be confused by the relativistic Bohmian mechanics [19], which interprets the hydrodynamic-like ve-locity as the real velocity of the quantum particle. Since the relativistic motion is always very fast, one would expect that far from equilibrium the friction force will become also nonlinear. In this case one can model it by a cubic friction, for instance, and derive a nonlinear diffusion equation analogous to Eq. (12) [20]

$$\partial_t \rho = \nabla \cdot \{\rho \sqrt[3]{\nabla[U + (1 - Q[\rho^{1/2}Q]/2mc^2)Q + k_B T \ln \rho]/b_3}\} \tag{14}$$

where $b_3$ is the specific friction constant of the macroscopic cubic friction force $-b_3 V^3$. The latter reflects a nonlinear thermodynamic relaxation and should not be confused by the nonlinear microscopic friction added usually in the Langevin equation [21, 22]. For instance, the average macroscopic effect of a microscopic cubic friction force $-b_3 \dot{r}^3$ is not $-b_3 V^3$ due to nonlinearity.

The present approach is better than the existing framework of quantum Brownian motion in terms of accounting for the quantum nature of the Brownian particle. It does not include, however, quantum effects in the thermal bath and this is a challenge for further research. The well-established non-relativistic quantum Brownian theory fails to describe correctly a free quantum Brownian particle and thus the present results could not reduce to it without introducing mistakes [17]. The most powerful contemporary description of quantum Brownian motion is based on the Feynman-Vernon approach [11], which seems rigorous since it starts from the Schrodinger equation. However, this approach introduces ad hoc an initial separable density matrix, which contradicts to the presumed initial pure state of the whole system. As a result the Feynman-Vernon approach fails to describe the quantum Brownian motion at short times and the corresponding kinetic equations are strictly linear. They do not preserve the positivity requirement of the density matrix as well [23]. This reflects further in the corresponding linear quantum Smoluchowski equation and explains its shortcomings at short times as discussed before.

## Appendix

A short presentation of the Manev potential [16] is given below. In the case of the classical Newtonian gravity Eq. (1) reads

$$E = \sqrt{m^2 c^4 + c^2 p^2} + U_N = Mc^2 - mGm_0/r \tag{15}$$

where $G$ is the universal gravitational constant and $m_0$ is a referent static mass. To make Eq. (15) consistent with the general relativity one should replace the particle mass $m$ in the Newtonian gravitational potential by the relativistic mass $M$. The latter depends, however, from the particle momentum and for this reason we should approximately express it as a function of the particle coordinate as follows

$$M = \sqrt{m^2 + p^2/c^2} \approx m\sqrt{1 + 2Gm_0/c^2 r} \approx m(1 + Gm_0/c^2 r + \cdots) \qquad (16)$$

Replacing now the gravitational mass in Eq. (15) by Eq. (16) yields

$$E = Mc^2 - mGm_0/r - m(Gm_0/cr)^2 = \sqrt{m^2 c^4 + c^2 p^2} + U_N + U_M \qquad (17)$$

where $U_M \equiv -m(Gm_0/cr)^2$ is the Manev semiclassical potential [24], which is the first relativistic correction to the classical Newton potential $U_N$. As a matter of fact, Newton has already proposed such an addition $C/r^2$ to his gravity potential [25], without knowing that it is due to the general relativity. Thus, he has explained the planet orbital precession, which is nowadays known to be due to the Einstein general relativity [26]. Manev is the first who gave the right physical meaning of this empirical potential, being known for more than three centuries.